\documentclass[aps,prl,floatfix,twocolumn,superscriptaddress]{revtex4}

\usepackage{graphicx}

\begin{document}

\title{Magnetovolume effect, macroscopic hysteresis and moment collapse in the paramagnetic state of cubic MnGe under pressure}

\author{N. Martin}\affiliation{Laboratoire L\'eon Brillouin, CEA, CNRS, Universit\'e Paris-Saclay, CEA Saclay 91191 Gif-sur-Yvette, France}
\author{I. Mirebeau}\affiliation{Laboratoire L\'eon Brillouin, CEA, CNRS, Universit\'e Paris-Saclay, CEA Saclay 91191 Gif-sur-Yvette, France}
\author{M. Deutsch}\affiliation{Synchrotron SOLEIL, L'Orme des Merisiers, Saint-Aubin, 91192 Gif-sur-Yvette, France}
\affiliation{Universit\'e de Lorraine, CRM2, UMR UL-CNRS 7036, BP 70239, 54506 Vandoeuvre-l\`es-Nancy, France}
\author{J.-P. Iti\'e}\affiliation{Synchrotron SOLEIL, L'Orme des Merisiers, Saint-Aubin, 91192 Gif-sur-Yvette, France}
\author{J.-P. Rueff}\affiliation{Synchrotron SOLEIL, L'Orme des Merisiers, Saint-Aubin, 91192 Gif-sur-Yvette, France}
\affiliation{Sorbonne Universit\'es, UPMC Univ Paris 06, CNRS, Laboratoire de Chimie Physique-Mati\`ere et Rayonnement, 75005 Paris, France}
\author{U.K. R\"ossler}\affiliation{IFW Dresden, PO Box 270116, 01171 Dresden, Germany}
\author{K. Koepernik}\affiliation{IFW Dresden, PO Box 270116, 01171 Dresden, Germany}
\author{L.N. Fomicheva}\affiliation{Vereshchagin Institute for High Pressure Physics, Russian Academy of Science, 142190 Troitsk, Moscow, Russia}
\author{A.V. Tsvyashchenko}\affiliation{Vereshchagin Institute for High Pressure Physics, Russian Academy of Science, 142190 Troitsk, Moscow, Russia}\affiliation{Skobeltsyn Institute of Nuclear Physics, MSU, Vorobevy Gory 1/2, 119991 Moscow, Russia}

\date{\today}

\pacs{}
\keywords{}

\begin{abstract}
Itinerant magnets generally exhibit pressure induced transitions towards non magnetic states. 
Using synchrotron based X-ray diffraction and emission spectroscopy, the evolution of the 
lattice and spin moment in the chiral magnet MnGe was investigated in the {\textit{paramagnetic}} 
state and under pressures up to 38\,GPa.
The collapse of spin-moment takes place in two steps. 
A first-order transition with a huge hysteresis around 7\,GPa transforms the system 
from the high-spin at ambient pressure to a low-spin state. The coexistence of 
spin-states and observation of  history-depending irreversibility 
is explained as effect of long-range elastic strains mediated by 
magnetovolume coupling. Only in a second transition, at about 23\,GPa, the spin-moment collapses.
\end{abstract}

\maketitle

Magnetism in electronic systems is fundamentally 
unstable with respect to lattice compression. 
Spin-state instabilities and transitions between
different band-magnetic states cause thermodynamic anomalies 
under temperature or pressure changes.
Probably, the best known anomaly of this type is the invar effect, 
yielding a paused thermal expansion around room-temperature in Fe-Ni alloys  
and various metallic 
materials  \cite{guillaume1897,Wassermann1990INVAR,Shiga1994}.
This effect is widely employed in industrial applications.
Generally, it is believed that modifications
in the magnetic behavior and magnetovolume coupling 
underly such anomalies.
An early explanation by Weiss has been 
based on existence of a high spin (HS) state with large volume
and a metastable low spin (LS) state of reduced volume. 
Thermal activation of the LS state 
counteracts the usual expansion of the lattice with 
increasing temperature \cite{weiss1963origin}.
Band-theory calculations on Fe-based alloys and compounds
with invar-anomalies later supported the basic assumption
of a discontinuous transition and the magneto-volume effect
as source of invar anomalies \cite{Moruzzi1986}. 
Similar effects with intermediate spin-states or reduced magnetic moments 
have been described in other solid-state systems, such 
as certain transition metal oxides \cite{raccah1967first,podlesnyak2006spin} 
or molecular complexes \cite{gutlich2004spin,Slimani2013}.
For the metallic invar-like systems,
a coherent physical picture of such magneto-volume effects and the fundamental
mechanisms could not be achieved.
Especially the existence of intermediate 
spin-states and discontinuous transitions between magnetic 
states is debated \cite{vanSchilfgaarde1999origin,brown2001temperature,matsumoto2011noncollinear},
while being suggested by several experiments 
\cite{abdelmeguid1989observation,odin1999magnetic,dubrovinsky2001pressure,rueff2001magnetism,decremps2004abrupt,yokoyama2011anharmonicity}.

MnGe belongs to the family of cubic chiral helimagnets, where the dominant ferromagnetism competes with spin-orbit coupling resulting in long wavelength helical spin-structures. Among these compounds, MnGe stands out. Its B20-structure (space group P2$_{1}$3) is metastable at room temperature and powder samples are obtained by high temperature (800-2200\,K) and high pressure (2-8\, GPa) quench during the synthesis. Interestingly, MnGe displays the shortest helical pitch ($\sim$ 30 \AA) of the B20 family \cite{Kanazawa2011,Makarova2012,Grigoriev2013,gayles2015dzyaloshinskii}, resulting in a giant topological Hall effect. To explain it, 
a complex skyrmion lattice was postulated, but its existence down to T and H $\simeq$ 0 
is debated \cite{Nagaosa2013,Deutsch2014b,Viennois2015}. 
On the other hand, an inhomogeneous fluctuating chiral phase 
was observed over a very large temperature range \cite{Deutsch2014b}. 

Following a theoretical prediction \cite{Roessler2012}, pressure-induced collapse of magnetism in MnGe should take place in two steps between the equilibrium HS-state towards the zero-spin (ZS) state through an intermediate LS-state. Evidence for a HS to LS transition in MnGe
was indeed found by high pressure neutron diffraction \cite{Deutsch2014a}.
At low temperature, the ordered Mn moment decreases with increasing pressure 
in the HS state up to a critical pressure  $P_{\text{C1}} \simeq 6$~GPa, 
then remains constant, in excellent agreement with calculations on the transition
between HS and LS spin-state.  The N\'eel temperature $T_{\text{N}}$ 
was seen to reduce at a rate of $-14~\text{K}\cdot\text{GPa}^{-1}$.
At an extrapolated pressure $P_{\text{0}} \simeq$ 13 GPa the magnetic 
long-range order should vanish, but the pressure-collapse between the 
LS and ZS-state was not observed, yet. 

In this work, we report the observation of a clear first-order transition 
around 7\,GPa at room-temperature, far above the
magnetic ordering-temperature $T_{\text{N}} \simeq$ 170\,K in MnGe,
and the spin-collapse in the paramagnetic state 
is found at about 23~GPa.
The results demonstrate a discontinuous evolution 
and a co-existence of two microscopic spin-states 
in an ordered metallic compound. 
This remarkable invar-like
effect highlights the importance of long-range lattice strains in the spin-transition taking place in a chemically clean system. Such elastic strains mediated by magnetovolumic effect are crucial to explain the anomalous properties of MnGe.

\begin{center}
\begin{figure}[h]
\includegraphics[width=8cm]{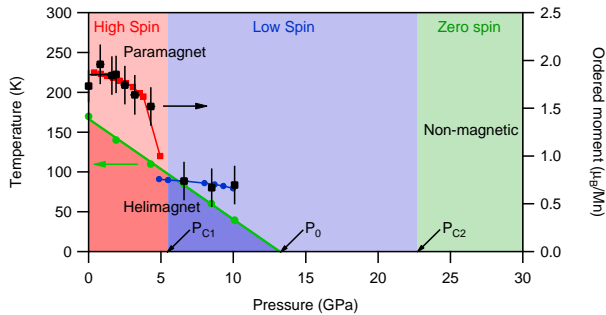}
\caption{\label{fig:Phase_diagram} (P,T)-phase diagram of MnGe inferred from neutron diffraction \cite{Deutsch2014a} and synchrotron-based x-ray techniques (this letter). Two well-separated regions with high ($0.7 \leq m \leq 1.8~\mu_{B}$, in red) and low ($m \sim 0.7~\mu_{B}$, in blue) ordered moment are clearly separated below the $T_{\text{N}}$(P) (green solid) line. The value of the experimentally determined ordered moment (black squares) along with the DFT rescaled results for the HS (red circles) and LS (blue circles) are also added.}
\end{figure}
\end{center}

 In order to understand the spin-state-transitions in MnGe,  we performed experiments 
 to detect the collapse of the {\it local} Mn moment, and  
not only the {\it ordered} one. 
 Moreover, by monitoring the evolution of the lattice parameter under pressure, we could detect magnetovolume effects induced by the different volumes and compressibilities of the HS and LS-state. Synchrotron-based X-ray techniques are ideally suited for these two tasks. Pressures well above P$_{\text{C,2}}$ can be reached by using membrane-type diamond anvil cells (DAC) with very small sample volumes. 
Fig. \ref{fig:Phase_diagram} is the $(P,T)$-phase diagram combining the earlier data
with results presented in this letter. 

X-ray powder diffraction (XRD) discriminates different spin states by measuring high-resolution $(P,V)$ equations of state (EoS). XRD was performed at room temperature on the PSICH\'E beamline of the synchrotron SOLEIL. No indication of a structural phase transition was found up to the highest pressure of 30~GPa, implying the absence of symmetry change or atomic displacements within the unit-cell (\cite{Supplement}). We therefore focus on the unit cell volume $V=a^{3}$ where $a$ is the cubic lattice constant deduced from Rietveld refinements. In order to describe its pressure-dependence, we use the {\it so-called} Murnaghan equation of state \cite{Murnaghan1944}, \(	V(P) / V_{0} = \left[1+P\, B_{\text{0}}'/B_{\text{0}}\right]^{-1/B_{0}'}\) where $B_{\text{0}}$ is the isothermal bulk modulus, $B_{\text{0}}'$ its first pressure derivative and $V_{\text{0}} = V(\text{P}\rightarrow 0)$. 

\begin{center}
\begin{figure}[h]
\includegraphics[width=8cm]{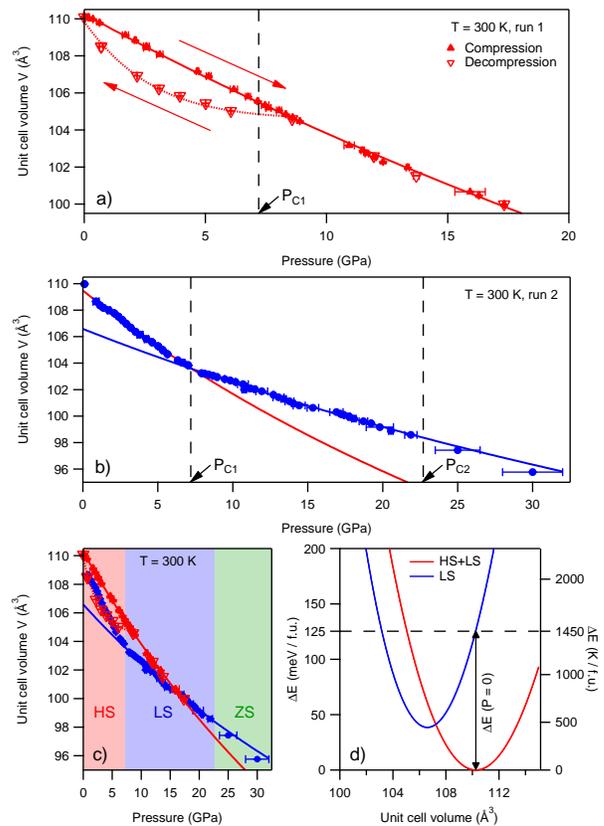}
\caption{\label{fig:EoS_lucia} Experimental $(P,V)$ EoS of MnGe deduced from our high pressure diffraction experiment. Results from the first a) and second b) run are overlayed in c). Solid lines are Murnaghan EoS fits to the data. d) Volume dependence of the energy for the HS+LS and LS states of MnGe calculated from the experimental parameters of the EoS.}
\end{figure}
\end{center}

In a first run, the applied pressure was increased up to $\sim$ 17 GPa - that is deeply inside the LS state - and then progressively released. The compression curve does not display drastic change of behavior (see Fig. \ref{fig:EoS_lucia}a). We attribute the EoS corresponding to this process to the initial HS-state which progressively transforms into the LS-state.
Parameters from Murnaghan EoS fit to the data are gathered in Tab. \ref{tab:EoS_DFT}. However, upon decompression, a remarkable structural hysteresis occurs, signaling the occurrence of a phase transition, across which a sizable LS proportion remains stabilized until pressure is fully released. 
In order to study the metastability of the HS-LS microstructure, also observed in invar alloys \cite{Gorria2009}, we have prepared a second sample that was loaded in a DAC at an initial pressure of $\simeq$ 7 GPa, maintained for about a week prior to the measurement. We have then quickly released the applied pressure and determined the EoS upon compression in the 0-30 GPa range. As seen on Fig. \ref{fig:EoS_lucia}b, a clear change in the EoS slope now occurs at {\it ca.} 7 GPa. A Murnaghan fit to the whole dataset gives unphysical values, $B_{\text{0}} = 90(5)~\text{GPa}$ and a very large $B_{\text{0}}' = 13.5(6)$
We conclude that the low pressure range concerns a HS-LS composition that depends on the thermal and pressure prehistory. Considering the pressure range above 7 GPa, we obtain the parameters for the EoS that can be attributed to the LS-state (Tab. \ref{tab:EoS_DFT}).

Our results show that a first-order transition takes place where specific volume and compressibility are discontinuously changed.
A maximal pressure of 7.2(5) GPa is estimated where the two states can coexist.
Following the observation on the evolution of the magnetic moments around the same pressure 
in the earlier magnetic neutron diffraction \cite{Deutsch2014a}, we identify the coexisting 
two states as HS and LS spin-states and the estimated maximum pressure compares rather well with the previous determination of P$_{\text{C1}}$.

\begin{table}[h]
\caption{\label{tab:EoS_DFT}Comparison of Murnaghan EoS parameters derived from DFT and determined experimentally.}
\begin{ruledtabular}
\begin{tabular}{lccc}
& $B_{\text{0}}$ (GPa) & $B_{\text{0}}'$ & $V_{\text{0}}$ (\AA$^{3}$)\\
\hline
HS+LS (run 1, Fig. \ref{fig:EoS_lucia}a) & $154(3)$ & $2.6(4)$ & $110.26(4)$\\
HS+LS (run 2, Fig. \ref{fig:EoS_lucia}b) & $119(7)$ & $3.4(9)$ & $109.48(8)$\\
LS (run 2, Fig. \ref{fig:EoS_lucia}b) & $237(3)$ & $4.3(2)$ & $106.6(4)$\\
HS (DFT) & $148$ & $ 2.5 $ & $107.9$\\
LS (DFT) & $165$ & $3.7 $ & $103.2$\\
ZS (DFT) & $177$ & $4.7 $ & $102.4$\\
\end{tabular}
\end{ruledtabular}
\end{table}

In Tab.~\ref{tab:EoS_DFT}, density functional theory (DFT) results on the $(P,V)$ equation of state (EoS) demonstrate the expected magnetovolume effects in the three different spin-states. This determination of the EoS calculation uses the full potential local orbital approach \cite{koepernik1999full}, and has an improved accuracy by using an extended set of basis states corresponding to the state-of-the-art \cite{Lejaeghere2015KSeos}. These calculations yield qualitatively similar changes 
for the EoS between HS and LS state (see Tab. \ref{tab:EoS_DFT}).
The lower equilibrium volumes $V_0$ can be explained by the fact that the DFT-results can
reproduce only the homogeneous $T=0$ ground-state (namely, they do not include thermal lattice expansion and also neglect certain effects of magnetic fluctuations).
There are notable differences for the bulk moduli $B_{\text{0}}$, but 
both findings agree in that the LS state possesses a smaller $V_{\text{0}}$ than the HS state, 
while being much less compressible.

There is a remarkable history dependence of the effective EoS 
and hence the spin state composition in MnGe. 
The results of lattice parameter {\it vs} pressure from both experiments runs 
are overlayed in Fig. \ref{fig:EoS_lucia}c.
The difference between the compression in the first and second run 
proves that the internal mixed state, starting at ambient pressure, 
must have been different. 
The two cycles probed here, by the mixed nature of initial states, 
clearly follow minor hysteresis loops. 

\begin{center}
\begin{figure}[h]
\includegraphics[width=8cm]{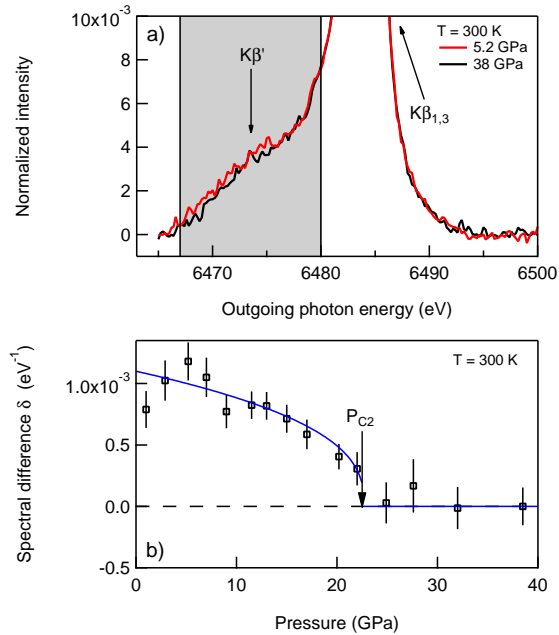}
\caption{\label{fig:XES_difference_P} a) Typical response measured at room temperature at 5.2 GPa (HS) and 38 GPa (ZS), illustrating the weak intensity decrease at the low energy satellite. Grayed zone denotes the energy range used for the integration. b) Pressure evolution of the integrated difference between spectra measured at each applied pressure value and the highest pressure spectrum. Solid line is a power-law fit (see text).}
\end{figure}
\end{center}

In order to address the magnetic collapse in MnGe on a local scale, we performed hard X-ray emission spectroscopy (XES) measurements under pressure at 300 \,K. Hard X-ray emission spectroscopy (XES) is sensitive to the local moment and earlier detected the pressure induced collapse of magnetism in invar alloys \cite{Rueff2001}.The emission spectra were recorded up to 38 \,GPa on the GALAXIES beam line of the synchrotron SOLEIL (see \cite{Rueff2015}). The element-specific photon emission at the $K\beta$ line of Mn is bound to spin-sensitive selection rules. 
While the system is excited by the incoming photons, the final state is characterized by a core hole 
($\underline{3p}$) that interacts with the $3d^{n}$ electrons via intra-atomic exchange. 
This results in the energy splitting of the emission line, yielding a main peak $K\beta_{1,3}$ at a photon energy of 6485 eV 
 paired with a low-energy shoulder $K\beta '$ located around 6475 eV (see Fig. \ref{fig:XES_difference_P}a). A decrease in the local Mn moment should yield a decrease in the intensity of the $K\beta '$ line relative 
to the main peak. In an itinerant magnet such as MnGe, the variation of the XES signal 
is however much smaller than in oxydes \cite{Mattila2007}. 
Phenomenologically, a way to monitor the evolution of the local moment is 
to consider the integral of the difference between a spectrum measured 
at a certain applied pressure $P$ with reference spectrum 
(in our case measured at $P = 38~\text{GPa}$), both being normalized 
to unity after appropriate background subtraction (\cite{Supplement}). The integral is solely performed 
around the satellite feature in order to get rid of pressure-dependent broadening of 
the main peak (\cite{Mao2014}). 

The result is displayed in 
Fig. \ref{fig:XES_difference_P}. The differential intensity $\delta$ decreases 
as pressure increases, up to about $\simeq 25~\text{GPa}$ where it saturates 
to 0 within error bars. This is indicative of another spin transition towards 
a state with a moment value that is lower than that of the LS state. 
Based on the good correspondence with DFT results \cite{Deutsch2014a}, 
we identify this transition as the local LS-ZS transition expected 
in this pressure range as a complete collapse of spin-polarization.
To estimate the associated critical pressure, we have fitted the data by the power law, 
\(\delta (P) =\delta_{\text{0}} \cdot \left(1-P/P_{\text{C2}}\right)^{\beta}\)  for \(P \leq P_{\text{C2}}\) 
and 0 otherwise,
yielding a critical pressure 
$P_{\text{C2}} = 22.7(1.8)~\text{GPa}$ with $\delta_{\text{0}} = 1.1(1) \cdot 10^{-3}$ and $\beta = 0.38(15)$.
Such a scaling is expected if there is a fluctuation-dominated transition 
from paramagnetic to non-magnetic state and should obey 3D-Ising criticality, but with pressure as
control parameter that drives the transition because of the different volumes of the LS and ZS-state. On the other hand, the HS-LS transition is hardly observable using the XES technique in this metallic compound, as the fine multiplet structure is not well established in comparison to the localized spin-states of an insulator
and may be influenced by temperature.

We also measured the evolution of the XES signal versus temperature 
in the range $5 \leq T \leq 300~\text{K}$ at ambient pressure (\cite{Supplement}). 
Essentially, the data are not indicative of a thermally driven HS-LS transition. Rather, a slight increase with temperature is seen that 
may be associated with a thermal re-population in the multiplet structure. 
At ambient pressure the Mn local moment at 300\,K is essentially 
in the same HS state as at low temperature in the ordered phase. This justifies {\it a posteriori} 
the X-ray experiments done at room-temperature and the comparison with  DFT data at 0\,K, 
but it raises the question why the HS-LS transition 
does not occur with temperature as in other spin crossover compounds.

In order to answer this question, we have calculated the energy curves of the LS and ambient pressure 
HS state from the relation $P = -\partial E / \partial V$ by using the Murnaghan EoS
(see Fig. \ref{fig:EoS_lucia}d) with the XRD results from run 1 for the HS-LS initial mixture and run 2 for the LS state (Tab. \ref{tab:EoS_DFT}).
One gets an energy gap $\Delta E \simeq 125~\text{meV}/\text{f.u.} \simeq 1450~\text{K}/\text{f.u.}$ between the two states at ambient pressure. 
This energy gap is larger than the temperatures where all reported magnetic measurements were performed (up to 300\,K typically). It explains why no HS-LS transition occurs versus temperature. At the same time, one can also speculate that the synthesis conditions (up to 2200\,K and 8 GPa) followed by a thermal quench could yield the nucleation of metastable LS  spin states in the dominant HS state. Such scenario would explain the large variability of magnetic properties reported in literature depending on the synthesis conditions of MnGe, which is not linked with impurities, off-stoechiometry or random disorder.

On the other hand, the suppression of LS-states in the ambient pressure HS-matrix,
and their metastable co-existence implied by the pressure-hysteresis 
requires a coupling that prevents a simple pressure-driven 
transition in a jump-like process. 
In the paramagnetic state, where long-range magnetic order is absent, the elasticity of the lattice remains the sole explanation for the realization of two energetically different spin-states in an extended pressure range.
A sizeable magnetovolume effect implies that 
the lattice is strained when locally a spin-state transition takes place.
These strains effectively mediate long-range couplings between the sites 
that slowly decay with distance as $r^{-3}$, acting 
 as an energetic barrier against a sizable nucleation of LS-states
 \cite{fratzl1999modeling,khomskii2004intersite,schwarz1995thermodynamics,schwarz2006thermodynamics}. Namely the local strains prevent the sites from permanently occupying the minority spin state. The spin state could be changed between HS and LS through thermal fluctuations, realizing the conditions of an "open" system. 
The elastic energy does not depend 
on the spatial arrangement of the LS-sites (a fact known as Crum-Bitter 
theorem for isotropic elastic two-phase bodies \cite{fratzl1999modeling}). 
In the ideal case of a homogeneous system, this barrier prevents 
the transformation until the stability limit of the matrix phase is reached. In the real case, the coexistence of the two spin states, considered as thermodynamic phases, occurs at a microscopic level, yielding hysteresis in the physical observables.
This  `thermodynamics of an open two-phase system' in a  coherent elastic solids 
has been analyzed in another context by Schwarz and Khachaturyan \cite{schwarz1995thermodynamics,schwarz2006thermodynamics}, 
but it exactly applies to the case of spin-state transitions because spin-states
can be changed by spin-lattice relaxation (\cite{Supplement}).
Improvements to this simple thermodynamic picture may introduce certain correlations
between sites of the nucleating phase, {\it e.g.} by the elastic anisotropy of the 
cubic lattice, but cannot fundamentally change this physical picture. 

In conclusion, the magnetic collapse in MnGe occurs in two-steps, in the paramagnetic regime as well as in the magnetically ordered state. The direct observation of the ultimate collapse ascertains the nature of the intermediate phase, which at low temperatures is a weak itinerant band ferromagnetic state. This somewhat contrasts with high pressure studies in other B20-helimagnets like MnSi and FeGe, where quantum phase transitions towards a non magnetic state have been found with intermediate regimes characterized by non Fermi liquid character and/or partial magnetic order \cite{Pfleiderer2004,Barla2015}. The huge  pressure-hysteresis at the transition between the ambient paramagnetic 
and the pressure-induced intermediate phase proves the co-existence of different spin-states.
The thermodynamic anomalies, in particular the  strong irreversibility marking the pressure-induced transformation in MnGe, can be explained by the long-range strains through the magnetovolume effect.
Anomalous non-equilibrium and transport behavior are also necessarily associated to
the magnetovolume effects, as observed in classical invar alloys.
Hence, the coexistence of spin-states, extending down to ambient pressure at room temperature 
in MnGe, should influence the anomalous helimagnetic fluctuations and transport properties of MnGe.

\begin{acknowledgments}
Experiments were carried out at SOLEIL synchrotron (Proposals No. 20140163 and 201440217). We warmly thank F. Baudelet, L. Nataf and P. Zerbino for experimental assistance. L.N.F. and A.V.T. acknowledge the Russian Foundation for Basic Research (Grant No. 14-02-00001). 
\end{acknowledgments}

\bibliographystyle{apsrev}
\bibliography{mnge_rx_v9,invar,phase_coexistence}

\section{Supplementary material}

\section{Samples}
Polycrystalline MnGe was synthesized at 8 GPa in a toroidal high-pressure apparatus by melting reaction with Mn and Ge. The purity of the constituents was 99.9\% and 99.999\% for Mn and Ge respectively. The pellets of well-mixed powdered constituents were placed in rock-salt pipe ampoules and then directly electrically heated to T $\simeq$ 1600$^\circ$C. The samples were subsequently quenched to room temperature before releasing the applied pressure as described by Tsvyashchenko\cite{Tsvyashchenko1984}. The sample quality was checked by X-ray and neutron diffraction, yielding an amount of impurity less than 2\%. The samples used in the experiments described in this letter were withdrawn from the very same synthesis that was used in previous studies\cite{Makarova2012,Deutsch2014a,Deutsch2014b}.

\section{X-ray powder diffraction}

\begin{center}
\begin{figure}[ht]
\includegraphics[width=8cm]{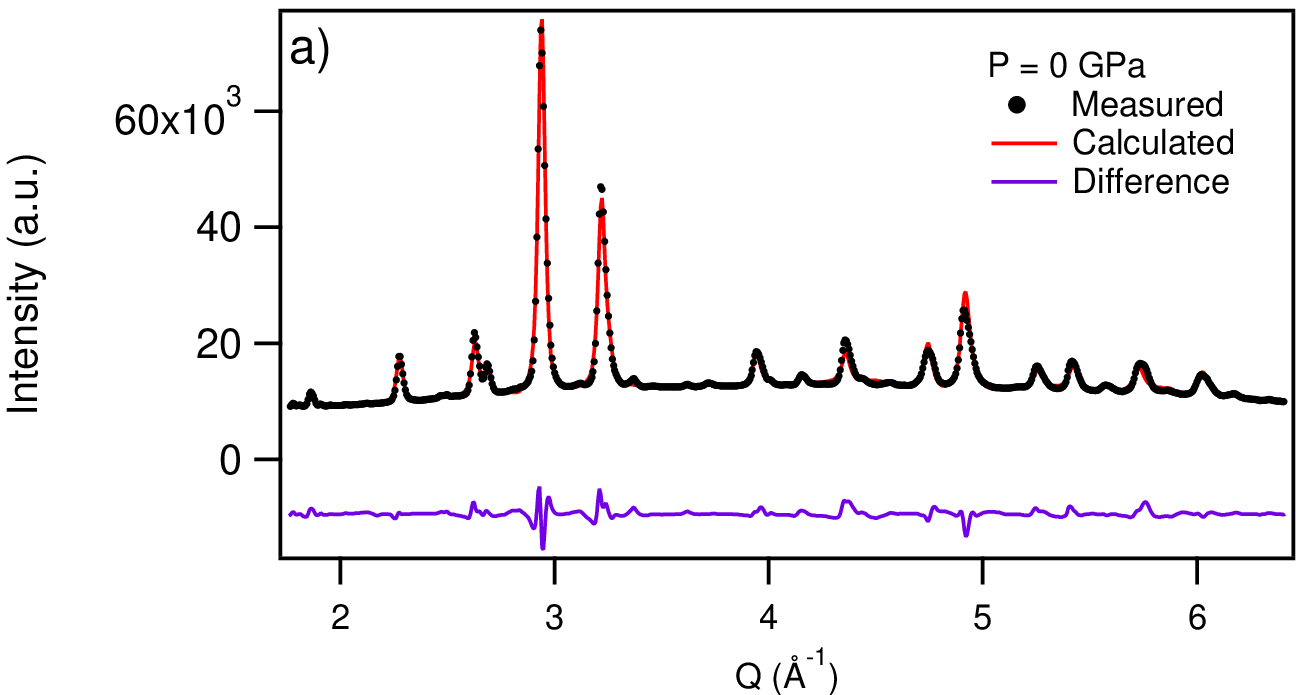}
\includegraphics[width=8cm]{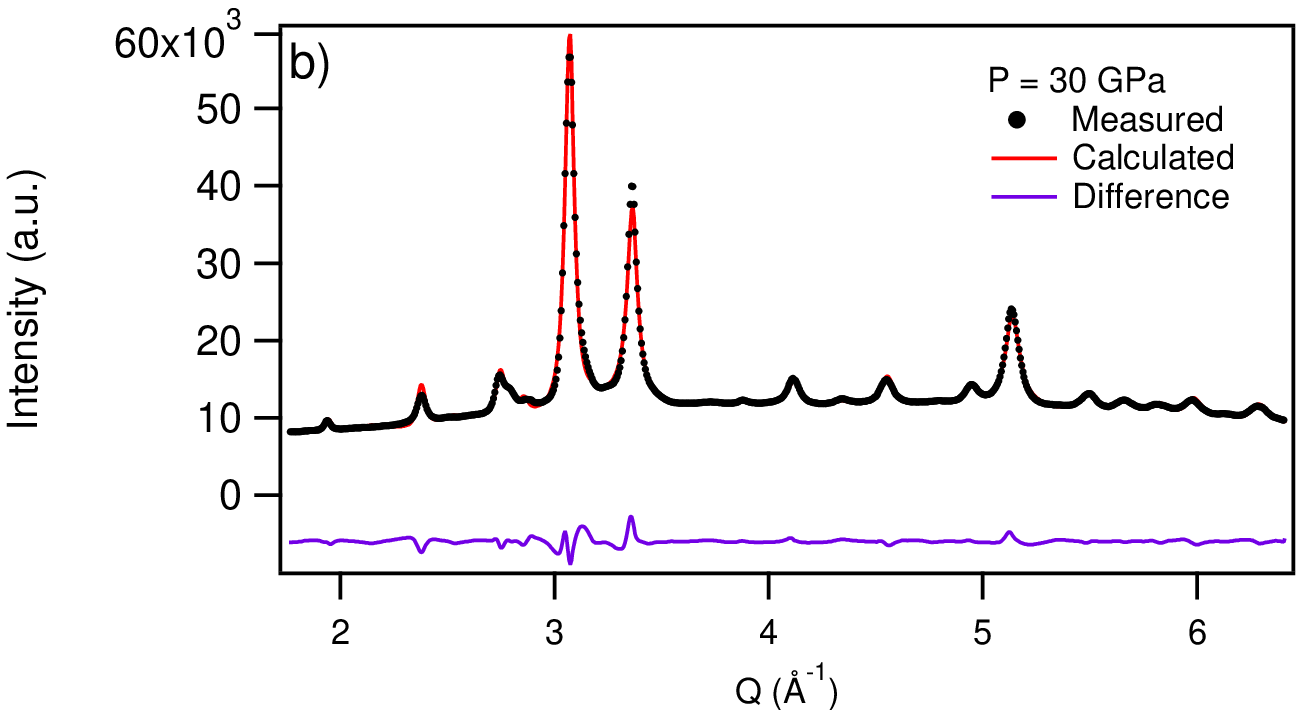}
\caption{\label{fig:diffractograms} Examples of measured diffractograms at a) 0 and b) 30 GPa, showing the high data quality. Red line is the result of a Rietveld refinement of the measured data.}
\end{figure}
\end{center}

High resolution x-ray powder diffraction (XRD) experiment was performed at the wiggler beamline PSICH\'E (Synchrotron SOLEIL). The incident X-ray wavelength was 0.3738 \AA.  Pressure was applied on the sample with the help of a diamond anvil cell (DAC). Ne was used as pressure transmitter as it offers excellent homogeneity. The X-ray patterns were averaged over Debye-Sherrer cones,  after supressing few non isotropic contributions from the pressure cell ({\it e.g.} Bragg spots from the diamond anvils). The X-ray patterns were refined using the Fullprof routine \cite{Fullprof1993}. The R$_{\text{F}}$ values situate between 1.3 and 3.0 $\%$. 
The very good refinements confirm that the sample remains in the B20 structure with negligible texture effect up to the highest pressure. Examples of refined diffractograms taken at low and high pressures are displayed in Fig. \ref{fig:diffractograms}. Besides the change in the lattice constant, the main effect of pressure is a Lorentzian peak broadening by a factor $\lesssim 2$, as seen in Fig. \ref{fig:diffractograms}.

\begin{center}
\begin{figure}[ht]
\includegraphics[width=8cm]{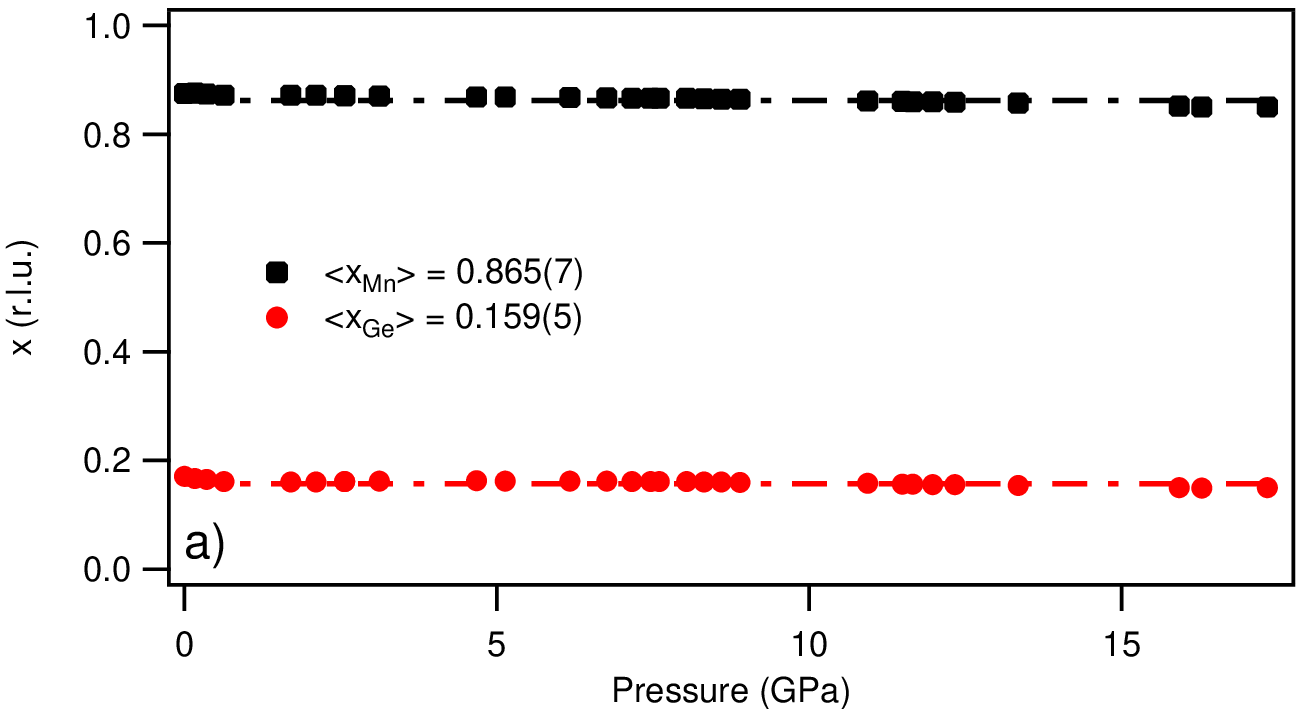}
\includegraphics[width=8cm]{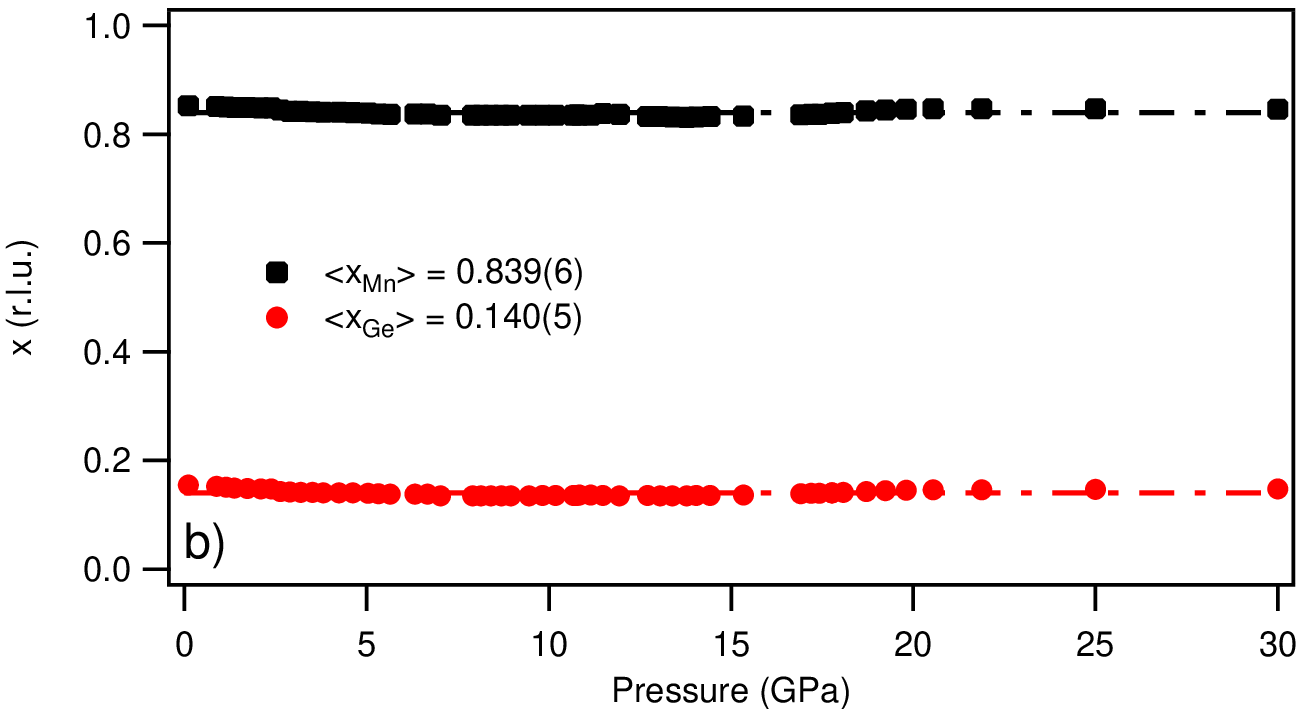}
\caption{\label{fig:xMnGe_Lucia} Pressure dependence of the positional parameters of Mn and Ge atoms -$x_{\text{Mn}}$ and $x_{\text{Ge}}$ respectively - as a function of the applied pressure for the first (a) and second (b) run (see main text).}
\end{figure}
\end{center}

As explained in the main text, no trace of a structural transition could be found throughout the data analysis. The positional $x$ parameters for Mn and Ge stay fairly constant and vary by less than 1 \% within the pressure range we explored (see Fig. \ref{fig:xMnGe_Lucia}). Such a variation most is most likely a bias of the refinement procedure, linked with anisotropic peak broadening occurring due to the geometry of the pressure cell.

\section{X-ray emission spectroscopy at the $K\beta$-line of Mn}

\begin{center}
\begin{figure}[ht]
\includegraphics[width=8cm]{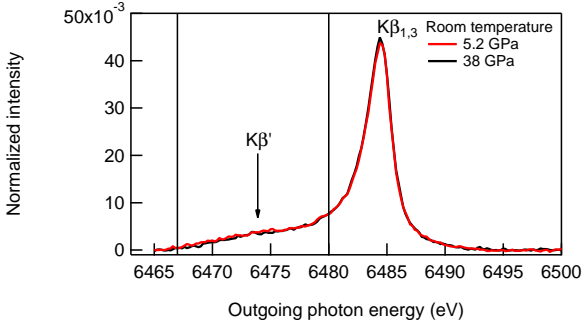}
\caption{\label{fig:xesxamples} Example of XES spectra measured at room temperature for P = 5 and 38 GPa. The vertical bars delimitate the integration boundaries.}
\end{figure}
\end{center}

Our x-ray emission spectroscopy (XES) measurements have been performed in a backscattering geometry ($2\theta = 135^{\circ}$) at the RIXS spectrometer of the undulator beamline GALAXIES\cite{Rueff2015} (Synchrotron SOLEIL). Emitted photon energies were determined by reflection on a Si(440) analyzer at 84$^{\circ}$ Bragg angle. Detection was ensured by an avalanche photodiode. We have used a DAC and a 4:1 methanol/ethanol mixture as pressure transmitter. The incoming photon energy was selected to be 7 keV, as a trade off between transmission through the pressure cell's diamond and emitted photon flux, while maximizing the signal-to-background ratio. An example of a full spectrum is displayed in Fig. \ref{fig:xesxamples}.

As explained in the main text, we have evaluated the evolution of the local Mn moment value as a function of pressure by means of a qualitative analysis. The latter is based on the comparison of the integral intensity of the $K\beta '$-low energy shoulder measured at a certain applied pressure $P$ with that of a reference. In our case, we assume the spectrum taken at $P_{\text{ref}} = 38~\text{GPa}$ to be representative of the non-magnetic state of MnGe. Thus, the integral difference that is linked with the local moment value \cite{Rueff1999,Rueff2001} is calculated as follows:

\begin{equation}
\begin{array}{r@{}l}
	\delta (P) &{}= \sum_{\omega_{\text{min}}}^{\omega_{\text{max}}} \left( \sigma (\omega,P) - \sigma (\omega,P_{\text{ref}}) \right)\\
	\sigma (\omega,P) &{}= \frac{I (\omega, P) - I_{\text{bg}} (\omega)}{\sum \left(I (\omega, P) - I_{\text{bg}} (\omega)\right)}
	\label{eq:xes_integraldiff}
\end{array}
\end{equation}

where $\omega$ is the emitted photon energy, $\omega_{\text{min}} = 6467~\text{eV}$, $\omega_{\text{max}} = 6480~\text{eV}$ and the subscript $\text{bg}$ refers to the linearly $\omega$-dependent background mainly originating from the high energy tail of the $K\alpha$ emission line. We have checked that changing the reference spectrum simply shifts the values of $\delta$ by a constant offset without altering its relative evolution. Moreover, extending the energy boundary to the whole measured range, we verify -as it must- that the $\delta$ turns out to be 0 at all pressures. This validates the background subtraction and normalization procedure.

\begin{center}
\begin{figure}[ht]
\includegraphics[width=8cm]{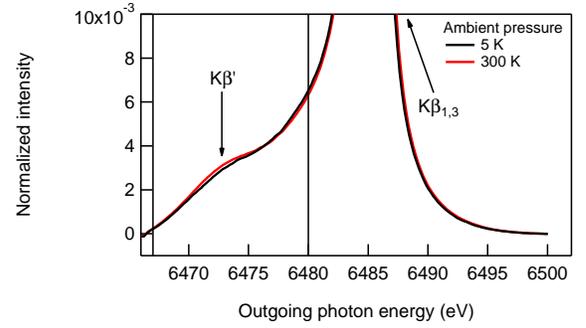}
\caption{\label{fig:XES_difference_T} Typical response measured at ambient pressure at 5 and 300 K, showing a small increase in the emitted intensity at the $K\beta'$ satellite position. One can note that the statistics is greatly improved compared to the high pressure study, as in the latter case the incoming and emitted photon flux was drastically reduced by the diamond anvils.}
\end{figure}
\end{center}

As mentioned in the main text, we have recorded some ambient pressure spectra between 5 and 300 K (see Fig. \ref{fig:XES_difference_T}). The spectral difference is {\it ca.} 5 times weaker between base and room temperature as compared with the value obtained under pressure between 0 and 38 GPa. Thus, this weak spectral change at the $K\beta'$ position doesn't support a HS-LS transition triggered by heating. Rather, it suggests a thermal population of the multiplet structure.

\section{High-pressure gauge}

\begin{center}
\begin{figure}[ht]
\includegraphics[width=8cm]{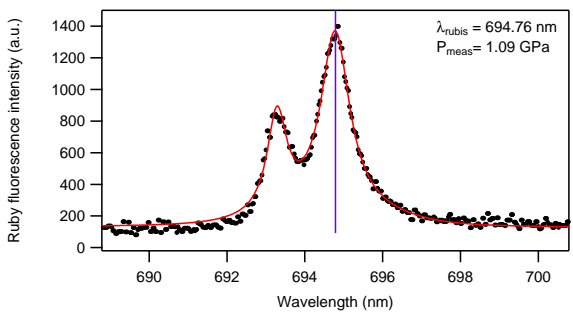}
\caption{\label{fig:joint} Picture of the sample within the DAC taken at the end of the XES experiment.}
\end{figure}
\end{center}

\begin{center}
\begin{figure}[ht]
\includegraphics[width=8cm]{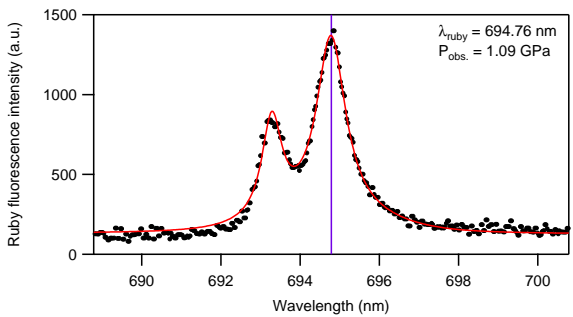}
\caption{\label{fig:fluospec} Example of ruby fluorescence spectrum as used for pressure determination.}
\end{figure}
\end{center}

In both the XRD and XES experiments, pressure was measured {\it in situ} before and after each single measurement via recording laser-stimulated ruby fluorescence lines. We have visually inspected the position of the ruby within the cell and checked that it was centered and embedded in the same medium as the sample (see Fig. \ref{fig:joint}). A double Lorentzian function was fitted to the obtained two-peaks spectra (see Fig. \ref{fig:fluospec}). The wavelength position $\lambda_{\text{m}}$ of the main line is then inputed into the following function
	
\begin{equation}
	P~\text{[GPa]} = \frac{A}{B} \cdot \left[\left(\frac{\lambda_{\text{m}} \text{[nm]}}{\lambda_{\text{0}}}\right)^{B}-1\right]
	\label{eq:ruby_fluo}
\end{equation}	
	
where $A = 1904~\text{GPa}$, $B = 7.715$\cite{}\cite{Zha2000} and $\lambda_{\text{0}} = 694.36~\text{nm}$, the latter being calibrated by measuring the response of a ruby at room pressure.

\pagebreak
\section{Electronic structure calculations by density-functional theory}
Theoretical calculations of the equation of state EoS have been done 
within the density-functional theory approach using the full-potential
local-orbital approach, as implemented in the FPLO code \cite{koepernik1999full}.
The generalized-gradient approximation (GGA) was used for
the exchange-correlation potential, and the lattice cell was
optimized for each calculated volume and the different spin-states.
The values of the Murnaghan EoS in table~I of the main text
have been determined from the theoretical (total) energy curves, $E_{tot}(V)$ 
in the range of volumes $98 < V < 111$~\AA$^3$.
For the LS-state, we apply the GGA*X corrections with reduction factor
for the XC-potential ($\xi=0.720$, as in Ref.~\onlinecite{Deutsch2014a}) 
in order to consider some of the long-range spin-fluctuations affecting 
the spin-state. At equilibrium volume 
the net spin moment is 0.8~$\mu_B/$~f.u. in this spin-state which 
is metastable within this calculation.
For the HS-state, the bare GGA results are reported which correspond 
to an electronic state that neglects all long-range spin-fluctuations ($\xi=1$). 
This ideal spin-state has a net spin-moment of  2.0~$\mu_B/$~f.u.
Some quantitative deviations in the EoS and magnetic properties 
between present evaluation and earlier calculations \cite{Roessler2012,Deutsch2014a}
are explained  by a refinement in the one-electron local-orbital basis.
For the present evaluation, we have used the doubled basis set
for valence band-states that was also used in a recent determination 
of the EoS of the crystalline states of the elements to be reported 
in the bench-marking comparing various DFT-codes performed and organized 
by LeJaeghere, Cottenier {\it et al.} \cite{Lejaeghere2015KSeos}$^{,}$\footnote{https://molmod.ugent.be/deltacodesdft}.
Using this improved basis set of the FPLO-code
yields maximum deviations of about 1~meV / ion for 
the energy curves near equilibrium states compared to other high-precision full-potential codes. 
In this respect, the solution of the EoS as given by the DFT Kohn-Sham equations 
are well converged.

\section{Spin-state transition in coherent elastic lattices}
In this section, we discuss in more details the elementary 
thermodynamic picture of the transition between different spin-states 
that we have used to identify long-range elastic stresses as 
the crucial factor for the large and history-depending hysteresis
in the high pressure XRD experiment.
Schwarz and Khachaturyan disscussed the co-existence of 
two solid phases $\alpha$ and $\beta$ 
under hydrogen-loading when only an 
external partial hydrogen pressure controls the amount
of interstitially disolved H-ions in the two lattices\cite{schwarz1995thermodynamics,schwarz2006thermodynamics}.
The thermodynamics of a paramagnet with a HS to LS 
transition with a magnetovolume-induced strains 
can be mapped onto the elementary model 
developed by them to explain the macroscopic hysteresis in such open two-phase systems.
The elementary model used by them to explain the
macroscopic hysteresis in such open two-phase systems
can be mapped onto the case of a paramagnet with
HS and LS transition with magnetovolume-induced strains.
Here, the spin-polarization of the electronic structure can 
be changed by spin-lattice relaxation, which means that 
the system can change the local spin-polarization and the
associated volume strain freely. In this sense 
we are discussing  an \textit{open} thermodynamic system
where two different phases can coexist, but the nucleation
causes long-range strains to appear in the system.
For simplicity only two spin-states $\nu=\{\text{HS},\text{LS}\}$ are considered
that can exist in a certain range of values
for the squared normalized spin-polarization given by the ratio $\mu_{\nu}=(M_{\nu}/M_{\nu}^{0})^2 \left( 0 < \mu_{\nu} < 1 \right)$, 
where $M_{\nu}$ is the actual local magnetization or spin-density
per volume and $M_{\nu}^{0}$ the full spin-polarization for the given spin-state.
Identifying the reduced $\mu_{\nu}$ 
with the H-concentrations $c_{\alpha}$ and $c_{\beta}$,
the elementary model can be formulated by simply 
rewriting the basic equations for free energy 
contributions of the different phases and the elastic strain.
In the following, we reproduce the equations and basic arguments from 
for such a paramagnet with magnetoelastic coupling using
the notation from Ref.~\onlinecite{schwarz1995thermodynamics,schwarz2006thermodynamics}
to mark the essential equivalence of this model with the 
case of hydrogen loading in a two-phase material.

The strain caused by the magnetovolume effects for the nucleation of sites 
with different spin-state in the matrix increases the elastic energy. 
The situation corresponds to the inclusion of misfitting spheres in the holes of a recipient elastic matrix. 
This energy cost reads: 
\begin{equation}
	E_{el} = N \, A \, \bar{\mu} \, (1-\bar{\mu}) \quad \text{with} \quad A = v_0 \, G_s \, \frac{1+\sigma}{1-\sigma} \, \epsilon_0^2 \quad ,
	\label{eq:elastic_energy}
\end{equation}
where $N$ is the number of lattice sites involved in the transformation, $v_{0}$ the volume of a single site, $G_s$ the shear modulus, $\sigma$ the Poisson ratio, $\epsilon_{0} = da/ad\bar{\mu}$ the volume dependence of the average spin moment and the average moment $\bar{\mu}$ is defined as: 
\begin{equation}
	\bar{\mu} = \omega \mu_{\text{LS}} + (1-\omega) \, \mu_{\text{HS}} \quad ,
	\label{eq:mu_bar}
\end{equation} 
where $\omega$ is the volume fraction of the LS phase.
Importantly, the form of Eq. \ref{eq:elastic_energy} is independent of the arrangement of the sites with deviating
spin-moment. 
In the extreme case, we may even have single lattice sites
undergoing HS to LS transitions in a HS-matrix.
Using Eq. \ref{eq:elastic_energy}, we can write the Helmholtz free energy per lattice site for a given phase:
\begin{equation}
	F_{\nu} (V,T,\mu_{\nu}) = f_{\nu} (V,T,\mu_{\nu}) + A_{\nu} \, \mu_{\nu} \, (1-\mu_{\nu}) \quad ,
	\label{eq:F_singlephase}
\end{equation} 
where the first term corresponds to the magnetic contribution and the second term expresses the magneto-elastic coupling.
To simplify the discussion of the basic mechanism, the elastic and materials coefficients are assumed to be the same in the two phases ($A_{\text{HS}} = A_{\text{LS}} = A$).
It is then possible to derive the expression for the total free energy of a two phases {\it closed} system via Eqs. \ref{eq:mu_bar}-\ref{eq:F_singlephase}:
\begin{eqnarray}
	\nonumber
	F_{\nu} (V,T,\bar{\mu},\mu_{\text{HS}},\mu_{\text{LS}},\omega) &=& \omega \, f_{\text{LS}} (V,T,\mu_{\text{LS}})\\
	&+& (1-\omega) \, f_{\text{HS}} (V,T,\mu_{\text{HS}})\\
	\nonumber
	&+& A \, \bar{\mu} \, (1-\bar{\mu})  
	\label{eq:F_twophases}
\end{eqnarray} 
In an {\it open} system, the average moment $\bar{\mu}$ is not fixed anymore and can be tuned by an appropriate external potential (in our case, the applied pressure). The corresponding Gibbs free energy reads:
\begin{eqnarray}
	\nonumber
	G_{\nu} (V,T,P,\bar{\mu},\mu_{\text{HS}},\mu_{\text{LS}},\omega) &=& F_{\nu} (V,T,\bar{\mu},\mu_{\text{HS}},\mu_{\text{LS}},\omega)\\
	&-& \kappa \, P \, \bar{\mu}
	\label{eq:G_twophases}
\end{eqnarray} 
where $P$ is the applied pressure and $\kappa$ a magneto-elastic coupling constant (in units of volume per moment) such as the average unit cell volume $V = \kappa \, \bar{\mu}$.
By virtue of Eq. \ref{eq:mu_bar}, Eq. \ref{eq:G_twophases} can be rewritten in the form of a second order polynomial function:
\begin{eqnarray}
	\nonumber
	G_{\nu} (V,T,p,\mu_{\text{HS}},\mu_{\text{LS}},\omega) &=& \phi_{0}\left(\mu_{\text{HS}}\right)\\
	&+& \phi_{1}\left(\mu_{\text{HS}},\mu_{\text{LS}}\right) \, \omega\\
	\nonumber
	&-& \phi_{2}\left(\mu_{\text{HS}},\mu_{\text{LS}}\right) \, \omega^{2} \quad ,
	\label{eq:G_twophases_polynom}
\end{eqnarray}
with
\begin{equation}
\begin{array}{r@{}l}
	\phi_{0} &{}= f_{\text{HS}}\left(\mu_{\text{HS}}\right) + A \, \mu_{\text{HS}} \, (1-\mu_{\text{HS}}) - \kappa \, P \, \mu_{\text{HS}}\\
	\phi_{1} &{}= (\mu_{\text{LS}}-\mu_{\text{HS}}) \, \left[\frac{f_{\text{LS}}(\mu_{\text{LS}})-f_{\text{HS}}(\mu_{\text{HS}})}{\mu_{\text{LS}}-\mu_{\text{HS}}} + A \, (1-2\mu_{\text{HS}}) - \kappa \, P\right]\\
	\phi_{2} &{}= A \, (\mu_{\text{LS}}-\mu_{\text{HS}})^{2}
	\label{eq:G_twophases_polynomialterms}
\end{array}
\end{equation}
Depending on the relative values of the $\phi_{i}$ ($i=\{0,1,2\}$) terms, the Gibbs free energy may monotonically increase or decrease as a function of the phase fraction $\omega$, yielding a minimum at $\omega = 0$ (macroscopic HS-state) or at $\omega = 1$ (macroscopic LS-state). An interesting third possibility arises when a maximum of $G$ occurs at an intermediate value $\omega^{*} = \phi_{1}/(2 \, \phi_{2})$ (see Fig. 1 of Ref. \onlinecite{schwarz1995thermodynamics}). In this case, the elastic strains create a macroscopic barrier against nucleation of a stable minority phase. For the spin transition to be triggered, the Gibbs free energy at $\omega = 0$ must cease to be a local minimum and hence, the linear term $\phi_{1}$ in Eq. \ref{eq:G_twophases_polynom} must be cancelled since $\phi_{2}$ is always $> \, 0$:
\begin{equation}
	\frac{f_{\text{LS}}(\mu_{\text{LS}})-f_{\text{HS}}(\mu_{\text{HS}})}{\mu_{\text{LS}}-\mu_{\text{HS}}} + A \, (1-2\mu_{\text{HS}}) - \kappa \, P = 0
	\label{eq:phi1_cancellation_0}
\end{equation}
The pressure at which the HS phase is stabilized can be calculated by differentiating Eq. \ref{eq:G_twophases} with respect to $\mu$:
\begin{equation}
	\kappa \, P \left(\mu_{\text{HS}}\right) = \left.\frac{\partial F_{\text{HS}}}{\partial \mu}\right|_{\mu=\mu_{\text{HS}}} = \left.\frac{\partial f_{\text{HS}}}{\partial \mu}\right|_{\mu=\mu_{\text{HS}}} + A \, (1-2\mu_{\text{HS}})
	\label{eq:dF_dmu_HS}
\end{equation}
Inserting Eq. \ref{eq:dF_dmu_HS} into Eq. \ref{eq:phi1_cancellation_0}, one eventually gets: 
\begin{equation}
	\frac{f_{\text{LS}}(\mu_{\text{LS}})-f_{\text{HS}}(\mu_{\text{HS}})}{\mu_{\text{LS}}-\mu_{\text{HS}}} - \frac{\partial f_{\text{HS}}(\mu_{\text{HS}})}{\partial \mu_{\text{HS}}} = 0
	\label{eq:phi1_cancellation_1}
\end{equation}
In this sense, the {\it stability limit} for the HS phase will be reached when the energy curves for the HS and LS states will have a common tangent for the first time, {\it i.e.} when the energy barrier for the stabilization of the LS phase is overcome under pressure. This reasoning may be applied to the backward LS-HS transformation. Since for the LS state:
\begin{equation}
	\kappa \, P \left(\mu_{\text{LS}}\right) = \left.\frac{\partial F_{\text{LS}}}{\partial \mu}\right|_{\mu=\mu_{\text{LS}}} = \left.\frac{\partial f_{\text{LS}}}{\partial \mu}\right|_{\mu=\mu_{\text{LS}}} + A \, (1-2\mu_{\text{LS}})
	\label{eq:dF_dmu_LS}
\end{equation}
we anticipate the observation of a finite square-like hysteresis of width $\kappa \, \left|P \left(\mu_{\text{HS}}\right)-P \left(\mu_{\text{LS}}\right)\right|$ in the $(P,V)$ equation of state of the material since $V$ depends on the average spin state.

Transposing these considerations to MnGe, it means that transformation between the HS and LS state will be marked by a large hysteresis of the unit cell volume through macroscopic magneto-elastic coupling, as observed in the first run of our XRD measurement (see main text). However, internal stresses like defects (originating from the high pressure synthesis of MnGe powder), but also crystallite sizes and their shapes, will massively influence the transformation processes and may favor the stable inclusion of minority LS sites in the ambient majority HS matrix. 
Thus, the HS-LS transition will not display a marked "jump-like" behavior at the critical pressure and will be replaced by a smooth crossover. An analogous case of macroscopic hysteresis phenomena and their dependence on microstructure are classical ferromagnets with dipolar stray-fields, as realized in the huge variability of
the hysteresis in permanent magnetic materials, where intrinsic magnetic properties are not
changed, but magnetization processes may yield differences in the coercive fields by orders
of magnitude. In the present discussion, for the sake of simplicity, we have neglected the role of magnetic correlations which are still sizable at room temperature, as seen {\it e.g.} by neutron diffraction\cite{Deutsch2014a} and small-angle scattering\cite{Altynbaev2014}. Also, unavoidable deviation from perfect hydrostaticity in the pressure medium was neglected. Anyway, the model presented above explains the essential features of the $V(P)$ equation of state determined experimentally.

\end{document}